\documentclass[11pt]{article}
\usepackage[dvips]{graphicx}
\usepackage{epsfig}
\usepackage{latexsym}
\newcommand{\be}{\begin{equation}}
\newcommand{\ee}{\end{equation}}
\newcommand{\eps}{\varepsilon}

\newcommand{\GeV}{{\rm GeV}}
\begin{document}
\begin{center}
        {\Large\bf Determination of the mass of the $\Lambda \Lambda$ 
dibaryon\\[2mm] 
by the method of QCD sum rules}\\
\vskip 0.51cm

{\bf S. A. Larin, V. A. Matveev, A. A. Ovchinnikov, and A. A. Pivovarov}\\
\vskip 0.31cm
Institute of Nuclear Research, USSR Academy of Sciences\\
\vskip 0.31cm
Submitted 5 October 1985\\
\vskip 0.31cm
{Yad. Fiz. {\bf 44}, 1066-1074 (October 1986)}
\vskip 0.91cm

{\bf Abstract}
\end{center}
The method of QCD sum rules is used to calculate the mass of a
hypothetical stable dihyperon $H$ with the quantum numbers of two $\Lambda$
hyperons, whose existence was predicted by R.~L.~Jaffe
[Phys. Rev. Lett. {\bf 31}, 195, 617(E) (1977)] in the framework of the MIT
quark-bag model. Within the accuracy of the method of QCD sum rules,
which in the present case is $\sim 20\%$, the results obtained here agree
with those of Jaffe. However, the method of sum rules does not make it
possible to determine whether the mass of the dihyperon $H$ lies above
or below the $\Lambda \Lambda$ threshold (i.e., whether $H$ is stable).

\vskip 2cm

The theoretical investigation of multiquark states ($Q^n \bar Q^m$,
$n + m > 3$) and the experimental search for them may provide
important information about the properties of the interaction of
quarks at large distances. In particular, major efforts, both
theoretical~\cite{Jaffe,Matveev} 
and experimental~\cite{Yokosawa} have been directed at the study of
the dibaryon spectrum ($n = 6$, $m=0$).

In this connection, interest attaches to Jaffe's
prediction~\cite{Jaffe} that there exists a stable six-quark $s$-wave
state -- a dihyperon $H$, which is a singlet with respect to both
color 
and
flavor (with strangeness $-2$) and has $J^P = 0^+$ and mass 
2150~MeV. The quantum numbers of $H$ are  identical to the quantum numbers of
$\Lambda \Lambda$  [two $\Lambda(1115)$ hyperons], and its mass is
smaller than the sum of the masses of two $\Lambda$ hyperons. Therefore $H$
can decay only through the weak channel (consequently, it is stable
with respect to the strong interactions).

The prediction that a dihyperon exists was obtained in the framework
of the MIT quark-bag model~\cite{Chodos}.  It is important to test
this model prediction solely on the basis of the fundamental
principles of QCD. Such a test can be made by means of the method of
QCD sum rules, using either the technique of finite-energy sum
rules~\cite{Chetyrkin} or that of Borel sum rules~\cite{Shifman}.
The method of QCD sum rules is based on the fundamental principles
of QCD, and it has proved its effectiveness in calculations of the
masses of mesons~\cite{Shifman,Reinders,Krasnikov,Kras1} and
baryons~\cite{Kras1,Ioffe,Chung}.

In the present paper, assuming the existence of the above-mentioned
dihyperon, we calculate its mass (with the aim of testing the
prediction of Ref.~\cite{Yokosawa}) in the framework of the Borel
version of the method of QCD sum rules~\cite{Shifman}.

According to the method of QCD sum rules, to calculate the $H$ mass we
must consider the correlation function
\be
\Pi(q)=i\int d^4x e^{iqx}\langle 0|T\{h(x)h^+(0)|0\rangle .
\ee
Here $h(x)$ is a scalar local current with the quantum numbers of the
dihyperon $H$, constructed from six quark fields, such that its
projection onto the real dihyperon state $|H(p)\rangle$ 
(we are working with
the assumption that this state exists) is nonzero: 
\be
\langle 0|h(0)|H(p)\rangle =\lambda_H, \quad p^2=m_H^2 .
\ee
Since such a current $h(x)$ is not unique, the question of its  
optimal choice
immediately arises. We recall that the problem of choosing the current
already arose~\cite{Ioffe,Chung} 
in the case of baryons, when the currents are
constructed from three quark fields. In our case, when the current is
constructed instead from six quark fields, this problem is much more
complicated, since the number of independent currents with the given
quantum numbers is much greater. The treatment of the current $h(x)$ in
the most general form, i.e., in the form of a linear combination of
all the independent local operators with the quantum numbers of
$H$, is a
very cumbersome problem. Therefore we confine ourselves to the choice
of a few of the simplest currents with the required quantum numbers
and analyze the dependence of our results on the properties of the
studied currents. As in the case of mesons and baryons, we shall
construct the current $h(x)$ from quark fields without derivatives.  

We
shall consider only the following six simplest (from the point of view
of the calculation of the correlation function $\Pi$) currents with the
quantum numbers of the dihyperon $H$ ( we recall that $H$ is a singlet
with respect to both color and flavor):
\[
h_1(x)=\eps^{A_1A_2A_3}\eps^{A_4A_5A_6}\eps_{a_1a_2a_3}\eps_{a_4a_5a_6}
(\psi_1 C \psi_2)(\psi_3 C\gamma_5 \psi_4)(\psi_5 C \psi_6)\ ,
\]
\[
h_2(x)=\eps^{A_1A_2A_3}\eps^{A_4A_5A_6}\eps_{a_1a_2a_3}\eps_{a_4a_5a_6}
(\psi_1 C \gamma_5\psi_2)(\psi_3 C\gamma_5 \psi_4)
(\psi_5 C\gamma_5\psi_6)\ ,
\]
\[
h_3(x)=\eps^{A_1A_2A_3}\eps^{A_4A_5A_6}\eps_{a_1a_2a_3}\eps_{a_4a_5a_6}
(\psi_1 C  \gamma_\mu\gamma_5\psi_2)(\psi_3 C\gamma_5 \psi_4)
(\psi_5 C \gamma_\mu\gamma_5 \psi_6)\ ,
\]
\[
h_4(x)=\eps^{A_1A_2A_4}\eps^{A_3A_5A_6}\eps_{a_1a_2a_3}\eps_{a_4a_5a_6}
(\psi_1 C \gamma_5\psi_2)(\psi_3 C\gamma_5 \psi_4)
(\psi_5 C\gamma_5 \psi_6)\ ,
\]
\[
h_5(x)=\eps^{A_1A_2A_4}\eps^{A_3A_5A_6}\eps_{a_1a_2a_3}\eps_{a_4a_5a_6}
(\psi_1 C \gamma_5\psi_2)(\psi_3 C\gamma_5 \psi_4)
(\psi_5 C\gamma_5 \psi_6)\ ,
\]
\be
h_6(x)=h_5(x)-(1/3) h_2(x)\ ,
\ee
where $\psi_n=\psi_{A_n}^{a_n}(x)$ is a quark field, $a_n$
is the color index, $A_n$ is the flavor index, 
the spinor indices are omitted, and a summation over repeated indices 
is understood; $\eps^{A_1 A_2 A_3}$ and $\eps_{a_1a_2a_3}$ 
are antisymmetric tensors which are invariant with respect to the
flavor group $SU^f(3)$
and the color group $SU_c(3)$, respectively; $C$ are 
the matrices of charge conjugation. [The choice of the current 
$h_6(x)$ will be discussed below.]

In accordance with the method of QCD sum rules, 
we shall calculate the correlation functions
\be
\Pi_j(q)=i\int d^4x e^{iqx}\langle 0|T\{h_j(x)h_j^+(0)\}|0\rangle 
\ee
by means of Wilson's operator expansion, assuming that the vacuum 
expectation values of the local operators (the so-called condensates) 
are nonzero. The calculations must be performed in the region 
$- q^2\ge 1~{\rm GeV}^2$. In this region, on the one hand, 
the effective strong interaction constant $\alpha_s$ is 
already small enough to calculate the coefficient functions of the
operator 
expansion by perturbation theory; on the other hand, the
nonperturbative 
corrections due to the nonzero vacuum expectation values of the local
operators 
are already quite important.

Because of the great complexity of the calculations of $\Pi_j$, we shall
confine 
ourselves to a calculation in the zeroth order in $\alpha_s$.
The fact that this approximation may be sufficient for calculations of 
the physical quantities in the framework of the method used here is
confirmed 
by the practice of using the method, in particular by the calculation
of baryon 
masses. We calculate  $\Pi_j$ in the linear approximation in the strange 
quark mass $m_s$
and in the zeroth order in the masses of the $u$ and $d$ quarks.
The results of our calculations are as follows:
\be
\Pi_1=\Pi_2=\frac{1}{16}\Pi_3={\rm const}\cdot\Pi_4=
\ee
\[
=\left\{\frac{2}{35}\frac{(-q^2)^7}{7!}\ln\left(\frac{-q^2}{\mu^2}\right)
+\frac{2}{15}\frac{(-q^2)^4}{4!}\ln\left(\frac{-q^2}{\mu^2}\right)a^2
-\frac{1}{18}(-q^2)\ln\left(\frac{-q^2}{\mu^2}\right)a^4
\right.
\]
\[
\left.
+m_s\left[-\frac{4}{9}\frac{(-q^2)^5}{5!}\ln\left(\frac{-q^2}{\mu^2}\right)a
+\frac{4}{27}\frac{(-q^2)^2}{2!}\ln\left(\frac{-q^2}{\mu^2}\right)a^3
+\frac{7}{324}\frac{1}{(-q^2)}a^5
\right]\right\}\frac{64}{(16\pi^2)^5}\ ,
\]

\[
\Pi_5=
\left\{\frac{23}{35}\frac{(-q^2)^7}{7!}\ln\left(\frac{-q^2}{\mu^2}\right)
-\frac{121}{15}\frac{(-q^2)^4}{4!}\ln\left(\frac{-q^2}{\mu^2}\right)a^2
+\frac{121}{36}(-q^2)\ln\left(\frac{-q^2}{\mu^2}\right)a^4
\right.
\]
\[
\left.
+m_s\left[\frac{346}{45}\frac{(-q^2)^5}{5!}\ln\left(\frac{-q^2}{\mu^2}\right)a
-\frac{242}{27}\frac{(-q^2)^2}{2!}\ln\left(\frac{-q^2}{\mu^2}\right)a^3
+\frac{17}{648}\frac{a^5}{(-q^2)}
\right]\right\}\frac{64}{(16\pi^2)^5}\ ,
\]

\[
\Pi_6=
\left\{\frac{43}{63}\frac{(-q^2)^7}{7!}\ln\left(\frac{-q^2}{\mu^2}\right)
-\frac{1081}{135}\frac{(-q^2)^4}{4!}\ln\left(\frac{-q^2}{\mu^2}\right)a^2
+\frac{1081}{324}(-q^2)\ln\left(\frac{-q^2}{\mu^2}\right)a^4
\right.
\]
\[
\left.
+m_s\left[\frac{3034}{405}\frac{(-q^2)^5}{5!}\ln\left(\frac{-q^2}{\mu^2}\right)a
-\frac{2162}{243}\frac{(-q^2)^2}{2!}\ln\left(\frac{-q^2}{\mu^2}\right)a^3
+\frac{209}{5832}\frac{a^5}{(-q^2)}
\right]\right\}\frac{64}{(16\pi^2)^5}\ .
\]
Here  $a=|16\pi^2\langle \bar q q\rangle|$,  isotopic
invariance    gives
$\langle \bar u u\rangle=\langle \bar d d\rangle=
\langle \bar q q\rangle$, and we also assume 
$\langle \bar s s\rangle=\langle \bar q q\rangle$ for
the strange quark condensate, since our analysis has shown 
that allowance for the difference between 
$\langle \bar s s\rangle$ and $\langle \bar q q\rangle$ 
practically does not alter our final results. In the expressions
(5) we have retained only the terms which survive after 
application of a Borel transformation (see below); in 
the calculation of the vacuum expectation values of multiquark 
operators, we have made use of the hypothesis of 
factorization~\cite{Chodos}; the factor $1/(16\pi^2)^5$ 
corresponds to the fact that the 
perturbative contribution is represented by five-loop diagrams.

For each correlation function $\Pi_j$ 
we can write the dispersion relation (below, we shall drop the 
index $j$, since the arguments are valid for all $j$)
\be
\Pi(Q^2)=\int_0^\infty \frac{\rho(s)ds}{s+Q^2}-\mbox{subtractions}\ , 
\quad Q^2 = - q^2\ ,
\ee
where the spectral density 
$\rho(s) = (1/\pi){\rm Im} \Pi( - s - i 0)$ behaves as 
$s^7$ for large $s$, so that eight subtractions must be made in the
relation (6). 
Applying a Borel transformation (i.e., the differential operator
\[
\hat B=\lim_{Q^2,n\to\infty} \frac{Q^{2n}}{(n-1)!}
\left(-\frac{d}{dQ^2}\right)^n|_{Q^2/n=y}
\]
which makes the subtractions vanish) to both sides of Eq. (6), we
obtain 
a Borel sum rule of the well known form
\be
\label{borelPi}
\Pi^B(y)=\frac{1}{y}\int_0^\infty e^{-s/y}\rho(s)ds,
\ee
where $\Pi^B(y)$ can be obtained in our approximation from (5) by the
action of the operator $\hat B$:
\be
\hat B Q^{2k}\ln \left(\frac{Q^2}{\mu^2}\right)=(-1)^{k+1}k!y^k,
\quad \hat B \frac{1}{Q^{2k}}=\frac{1}{(k-1)!}\frac{1}{y^k}\ .
\ee

In order to calculate the mass of the dihyperon, it is necessary to
choose some 
phenomenological ansatz for the spectral density $\rho(s)$. The
simplest procedure 
which has proved to be successful in the calculation of meson and
baryon masses 
is to represent $\rho$ as the sum of the pole contribution of the unknown
lowest state of 
the given channel and an effective continuum beginning at some
effective threshold $W$ 
and approximating the contributions of all the higher states. We shall
also take 
a spectral density in the "pole plus continuum" form
\be
\label{spectraldens}
\rho(s) =\lambda_H^2\delta(s-m_H^2) +\theta(s-W^2) \rho^{cont}(s)\ ,
\quad \rho^{cont}(s)=(1/\pi){\rm Im} \Pi(-s-i0)
\ee
where the spectral density of the continuum is determined as the
imaginary part of the complete approximation calculated for $\Pi$.
Substituting the spectral density (\ref{spectraldens}) 
into Eq.~(\ref{borelPi}) and transferring
the continuum 
contribution to the left-hand side, we obtain a sum rule for the
correlation 
function $\Pi_1$,
\[
\frac{2}{35}y^7-\frac{2}{15}a^2 y^4-\frac{1}{18}a^4 y
+m_s\left(-\frac{4}{9}a y^5-\frac{4}{27}a^3 y^2
+\frac{7}{324}a^5\frac{1}{y}\right)
\]
\be
-\frac{1}{y}\int_{W^2}^\infty e^{-s/y}\tilde \rho^{cont}(s)ds=
\frac{1}{y}\tilde \lambda^2 e^{-m_H^2/y}
\ee
and a sum rule for the correlation function $\Pi_6$,
\[
\frac{43}{63}y^7+\frac{1081}{135}a^2 y^4+\frac{1081}{324}a^4 y
+m_s\left(\frac{3034}{405}a y^5+\frac{2162}{243}a^3 y^2
+\frac{209}{5832}a^5\frac{1}{y}\right)
\]
\be
\label{sumrule11}
-\frac{1}{y}\int_{W^2}^\infty e^{-s/y}\tilde \rho^{cont}(s)ds=
\frac{1}{y}\tilde \lambda^2 e^{-m_H^2/y}
\ee
where
\[
\tilde \rho^{cont}(s)=\frac{(16\pi^2)^5}{64} \rho^{cont}(s),
\quad \tilde \lambda^2=\frac{(16\pi^2)^5}{64} \lambda^2\ .
\]

The sum rules for $\Pi_{2,3,4}$ are identical to (10) apart from an
unessential overall factor. We have also not written down the sum rule
for the correlation function $\Pi_{5}$, since the coefficients in
$\Pi_{5}$  are numerically close to the coefficients in the
correlation function $\Pi_{6}$ [see Eq.~(5)], so that the results
obtained from the sum rule (11) for $\Pi_{6}$ are practically
identical to the results of $\Pi_{5}$.

The sum rules must be considered in an admissible interval of values
of $y$, in which the contributions of the higher power corrections and,
at the same time, the contribution of the effective continuum must not
be too large. We shall call this interval $\Omega$ and impose the following
restrictions for its determination:
1) the continuum contribution is less than 50\% of the calculated
approximation for $\Pi^B(y)$ which means that the continuum
contribution 
is
smaller than the contribution of the dihyperon $H$ to the sum rule;
2) the sum of all the nonperturbative corrections, excluding the
   leading nonperturbative term $\sim \langle \bar q q\rangle^2y^4$,
   is less than 50\% of the calculated approximation for $\Pi^B(y)$. 
(Although
   with respect to the dimensions of $y$ the term 
$\sim m_s\langle \bar q q\rangle^2 y^5$  
is the leading nonperturbative term, it is
   natural to include it in the constrained sum, since it is 
$\sim m_s$ and numerically the case $m_s = 0$ should not differ
   fundamentally from the case of broken $SU^f(3)$ symmetry, 
$m_s\ne 0$; moreover, if the term 
$\sim m_s\langle \bar q q\rangle^2 y^5$
   is not included in the constrained sum, this has little influence
   on the results.)

The choice of some {\it a priori} limit (in our case, 50\%) is highly
conditional. In the final anaysis, its correctness is substantiated 
{\it a posteriori} 
by the consistency of the sum rule in the interval $\Omega$ which
is found.

Before particularizing the procedure of fitting the sum rules, we
shall discuss the choice of the numerical value of the vacuum
expectation value $\langle \bar q q\rangle$, 
which depends on the normalization point $\mu$. We
shall choose $\mu^2\sim y \in\Omega$. Then the renormalization-group
factor 
$[\alpha_s(y)/\alpha_s(\mu^2)]^{\delta_n}$ 
in the coefficient function of a given operator $O_n$ 
(where $\delta_n$
is determined by the anomalous dimenions of $O_n$ and of the
current $h$) becomes $\sim 1$ for $y \in\Omega$. 
Therefore we shall neglect the
effects of the anomalous dimensions. In our case, the analysis of the
interval $\Omega$ shows that it is necessary to take
$\mu \sim 1~{\rm GeV}$. If we begin
with the value from Ref.~6, 
\[
|\langle \bar q q\rangle|_{\mu \sim 200~{\rm MeV}}=(240~{\rm MeV})^3
=0.014~{\rm GeV}^3
\]
then the application of the renormalization group gives the first estimate
$|\langle \bar q q\rangle|_{\mu \sim 1~{\rm GeV}} = 0.024~{\rm GeV}^3$.
On the other hand, if we recalculate the value of $\langle \bar q
q\rangle$  according to the PCAC formula  $\langle \bar q q\rangle  =
- (1/2)f_\pi^2 m_\pi^2/(m_u + m_d)$, then by using the value  $m_u +
m_d|_{\mu \sim 1~{\rm GeV}}= 16~{\rm MeV}$ which is now adopted for
the masses of the light quarks~\cite{Gasser} we obtain the second estimate
$|\langle \bar q q\rangle|_{\mu \sim 1~{\rm GeV}} =0.011~{\rm
GeV}^3$. It can be seen that those two values differ by a factor $\sim
2$.

Our final results depend little on the variation of 
$\langle \bar q q\rangle$ in the
indicated interval. This is perfectly natural. The value of $\langle
\bar q q\rangle$
is the only number which is built into our sum rule (inclusion of the
terms $\sim m_s$ should not substantially alter the
results). Therefore it follows from dimensional arguments that the
mass of the dihyperon is $m_H\sim (|\langle \bar q q\rangle|)^{1/3}$.
Consequently, $m_H$ should not change by more than
$\sim 30\%$ if $\langle \bar q q\rangle$ varies by a factor 2.

We must determine three parameters ($m_H, W,\lambda$) by fitting 
the sum rule in expression~(10) [or~(11)], i.e. on the basis of agreement 
between the left and right hand sides [which will be denoted by 
$L(W,y)$ and $R(m_H,y)$,
respectively] of the sum rule~(10) [or~(11)] in the admissible
interval $\Omega$. We rewrite the sum rule in the form
\be
L(W,y) y e^{m_H^2/y}=\tilde \lambda^2 .
\ee
Let $F(m_H,W,y)$ denote the left hand side of Eq.~(12). 
Since $\tilde\lambda^2$ must
not depend 
on $y$, we shall determine the values of $m_H$ and $W$ by requiring
the minimal 
dependence of $F$ on $y$ in the interval  $\Omega$. 
Thus, only two parameters $m_H$ and $W$
are effectively fitted, and $\tilde\lambda^2$ is determined simply as the
mean value 
$\bar F(m_H,W)$ of the function $F$ 
in $\Omega$ for the values of $m_H$ and $W$ which are found.
We introduce a criterion for the determination of $m_H$ and $W$. 
We consider the relative deviation of $F(m_H,W,y)$ from its mean value:
\be
\Delta=
\left|\frac{F(m_H,W,y)-\bar F(m_H,W)}{\bar F(m_H,W)}\right|=
\left|\frac{FL(W,y)-R(m_H,y)}{\bar R(m_H,y)}\right|\ .
\ee
We shall fix the values of $m_H$ and $W$ by requiring the largest
possible interval $\Omega$ (the upper limit of $\Omega$ depends on $W$)
subject to the condition that the relative error $\Delta$ does not
exceed a few percent (i.e., the function $F$ is practically a constant
in $\Omega$,  since it is a smooth function of $y$). For
definiteness, we require that $\Delta < 5\%$.

We analyzed the accuracy of the determination of the parameters on the
basis of the fitting procedure described above as follows. The
criterion for $\Delta$ was varied by a factor 2 in each direction from
the given value $\Delta  = 5\%$, and the value of the quark condensate
$\langle \bar q q\rangle$ was varied between the limits described
above. The corresponding changes in the values of $m_H$ and $W$ 
determined
from the fit were not more than  20\%. Therefore we can say that our
sum rule fixes $m_H$ and $W$  with 20\% accuracy. The parameter
$\lambda^2$ is determined with much lower accuracy with an accuracy up
to a factor 2. The greater accuracy in the determination of $m_H$ and
$W$ than in that of $\Delta$  is natural, since $m_H$ and $W$ appear
in the arguments of exponential functions and, consequently, the sum
rule is more sensitive to their variation.

Our analysis has shown that the sum rule~(10) for the currents 
$h_1- h_4$
is poorly fitted (in other words, it is not possible to satisfy the
criterion introduced above). At the same time, the sum rule (11) for
the current $h_6$ is well fitted over a
rather large interval $\Omega$.
In particular, 
for $\langle \bar q q\rangle = - (240~{\rm MeV})^3$ 
the results of fitting the sum rule (11) are as follows. 
For $m_s = 0$ we obtain
\be
m_H=2.0~\GeV,
\quad W=2.8~\GeV,
\quad \tilde \lambda^2=70~\GeV^8\ ,
\ee
and for $m_s =0.2~\GeV$ (Ref.~12)
\be
m_H=2.4~\GeV,
\quad W=3.2~\GeV,
\quad \tilde \lambda^2=150~\GeV^8\ .
\ee
The $y$ dependences of the left and right hand sides of the sum 
rule~(11) for the parameter values (14) and (15) which we have found are
shown in Figs. 1a and 1b, respectively.  

\begin{figure}[htb]\begin{center}
\vbox{\includegraphics[angle=-90,width=0.7\textwidth]{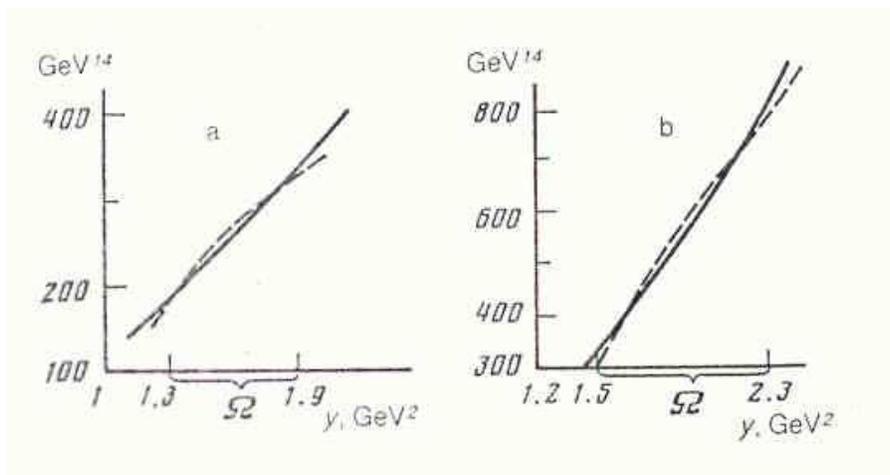}}
\caption{Dependence of the left hand side (continuous curves) and right
hand side (broken curves) of the sum rule~(11) on $y$ for the
parameter values in (14) and (15): a) $m_s=0$;
b) $m_s=0.2~\GeV$.}\end{center}
\end{figure}

We recall the values obtained by Jaffe~\cite{Jaffe} 
for the mass of the dihyperon $H$ (the values of the dibaryon
masses obtained in the quark-bag model correspond to poles of the 
$P$-matrix~\cite{Low} and the physical states have 
masses 100-200 MeV lower than
those predicted by this model): 
\be
\label{MIT-bag}
m_H|_{m_s=0}= 1.76~\GeV, \quad m_H|_{m_s\ne 0}=2.150~\GeV\ .
\ee
It can be seen that the values~(14,15) obtained for the dihyperon mass by
the method of QCD sum rules agree within the accuracy of this method
(in our case, the mass is determined with an accuracy ($\sim 20\%$) with the
results~(\ref{MIT-bag}) of the MIT quark-bag model.  

However, there immediately
arises the question of what makes the current $h_6$, for which the
results~(14,15) are obtained (the results for $h_5$ are practically
identical to them), preferred over the currents $h_1-h_4$ 
[see Eq.~(3)] which lead to the "poor" sum rule~(10).  

In our opinion, the answer is
that the current $h_6$ satisfies the following two "physical"
requirements. First, there exists a nonzero nonrelativistic
limit  for it [i.e. if the quark field $\psi(x)$ is represented 
in the standard
manner in terms of the small and large components, the term containing
only the large component will be nonzero]. On the other hand, the
currents $h_1$ and $h_4$ do not possess a nonrelativistic limit. We note
that in Refs.~\cite{Ioffe} and~\cite{IofShi} 
it was already pointed out that the existence
of a nonrelativistic limit is a desirable property for the
construction of currents in working with the method of QCD sum rules.

Second, we represent the current $h_6$ in the form of a product 
``singlet$\otimes$singlet'' with respect to color, i.e., in the 
form $h_6 =\Psi_1(x)\Gamma \Psi_2(x)$,
where each of the operators $\Psi_1$, and $\Psi_2$ 
is a color singlet and $\Gamma$ is some
combination of Dirac $\gamma$ matrices. Then we obtain the representation
(the currents $h_j$ are as a whole singlets with respect to both color
and flavor) 
\be
h_6(x)=(\Psi_{A_3}^{A_4}(x))^\alpha (C\gamma_5)_\alpha^\beta 
(\Psi_{A_4}^{A_3}(x))_\beta
\ee
where the color-singlet operator $\Psi$ is a flavor octet:
\be
\label{flavorOctet}
(\Psi_{A_3}^{A_4})_\alpha=
\eps_{a_1a_2a_3}(\psi_{A_1}^{a_1} C\gamma_5 \psi_{A_2}^{a_2})
(\psi_{B_1}^{a_3})_\alpha \eps^{A_1A_2B_2}
(\delta_{A_3}^{B_1}\delta_{B_2}^{A_4}
-(1/3)\delta_{B_2}^{B_1}\delta_{A_3}^{A_4})\ .
\ee
Here $\alpha$ and $\beta$ are
spinor indices, and the remaining notation is the same as in~(3).  

We
note that the flavor octet~(\ref{flavorOctet}) 
has the quantum numbers of the baryon
octet and has been used to calculate the characteristics of baryons by
the method of QCD sum rules~\cite{Ioffe,Chung}.
Thus, if the current $h_6$ is
represented in the form ``singlet$\otimes$singlet'' with respect to
color, it
has the structure ``octet$\otimes$octet'' with respect to flavor (each color
singlet is a flavor octet). Since the hyperon $\Lambda$ 
is a member of the
baryon flavor octet and the quantum numbers of the dihyperon $H$ are
identical to the quantum numbers of $\Lambda\Lambda$, 
it seems natural to construct
a flavor singlet $h(x)$ in the form  ``octet$\otimes$octet''.  

Physically, we can
imagine the following picture. If we split the colorless dihyperon $H$
into two colorless clusters and separate them to a large distance, we
obtain just two hyperons. However, if the currents $h_1-h_3$, are
represented in the form ``singlet$\otimes$singlet'' with respect to color, they
will also have the flavor structure ``singlet$\otimes$singlet'' (each color
singlet is also a flavor singlet), and therefore their structure seems
less natural than the structure of $h_6$.  

As a result, we conclude that
the current $h_6$ is more "physical" than the currents $h_1-h_4$, 
since it
has a nonrelativistic limit and is constructed as
``octet$\otimes$octet'' 
with
respect to flavor (in the sense indicated above). However, the
currents $h_1-h_4$ are "poor" because they either do not possess a
nonrelativistic limit (the currents $h_1$, and $h_4$) or have the flavor
structure ``singlet$\otimes$singlet'' (the currents $h_1,h_2$, and
$h_3$). 
The
current $h_5$ (we recall that the results for it are practically
identical to the results for $h_6$) has the flavor structure 
``nonet$\otimes$nonet'', and the ``octet$\otimes$octet'' 
term in it dominates over the ``singlet$\otimes$singlet'' term.  

We must make the reservation that formally the method
of QCD sum rules dictates {\it a priori} 
the only condition for the choice
of the current: it must possess the required quantum numbers. However,
{\it a posteriori} (after the fit) positive results were obtained only for
the current satisfying the two requirements formulated above. In our
opinion, this indicates the existence of a criterion which makes it
possible to select the optimal ("physical") current from the set of
currents with the quantum numbers of the given channel. We note that
this conclusion is confirmed also by the results of
Ref.~\cite{LarMat}, 
in which
the deuteron mass was calculated by the same method. In that paper,
the current was also constructed as ``baryon$\otimes$baryon'', 
and positive
results were obtained only when in the current the term having a
nonrelativistic limit dominated over the term not having this limit.

Unfortunately, the relatively low accuracy of our method in the
determination of the mass ($\sim 15\%$) does not make 
it possible
to draw a conclusion on the basis of our values~(14,15) about whether the
mass of the dihyperon $H$ lies below or above the $\Lambda \Lambda$ 
threshold (equal to 2.23~GeV). 
(We recall that the high accuracy of the MIT quark-bag
model permitted Jaffe to conclude that $H$ lies below the 
$\Lambda\Lambda$ 
threshold
and is therefore stable.) On the basis of our analysis, we can say
that the method of QCD sum rules agrees with the possible existence of
a dihyperon $H$ either below or above the $\Lambda \Lambda$ 
threshold (since in both
cases it is possible to represent the spectral density $\rho$ 
in the form
of the sum of a pole contribution from $H$ and a contribution from an
effective continuum).  

We shall consider in detail the case in which $H$
lies below the $\Lambda \Lambda$ threshold. 
In this cases, if the binding energy $\eps= 2m_\Lambda - m_H$ 
is sufficiently small (what "small" means will be
particularized below), there can occur the interesting effect of
resonance enhancement of the contribution of the $\Lambda \Lambda$ state of the
continuum~\cite{LarMat}. 
To describe this effect, let us consider the contribution
of the state $|\Lambda \Lambda\rangle$ 
(two free hyperons) to the spectral density: 
\be
\label{eq18}
\rho_{\Lambda \Lambda}(q^2)=(2\pi)^3
\langle 0|h(0)|\Lambda \Lambda\rangle
\langle \Lambda\Lambda|h^+(0)|0\rangle
\delta(q-p_1-p_2)
\ee
where $p_1^2=p_2^2= m_\Lambda^2$, and an integration over the phase space of 
the $\Lambda \Lambda$
state is understood. (We note the inclusion of the contributions of
the $\Sigma\Sigma$ and $\Xi N$ 
states in the approximation of exact $SU^f(3)$ symmetry
with allowance for only the flavor-singlet states does not alter the
results.) For the matrix element 
$\langle 0|h(0)|\Lambda \Lambda \rangle  =f_h(s)$, 
$s = q^2$, we
can write the dispersion relation
\be
\label{pole19}
f_h(s)=\frac{\lambda_H g_{H\Lambda}}{m_H^2-s}+
\int_{4m_\Lambda^2}^\infty\frac{\rho_h(s)ds'}{s'-s}-\mbox{subtractions}
\ee
where
$\lambda_H$ is defined in~(2) and $g_{H\Lambda}$ 
is the constant for the interaction of
the dihyperon $H$ with $\Lambda \Lambda$.  
On the other hand, for energies which are
not very large, when only the elastic channel of scattering of
$\Lambda$ by $\Lambda$
is possible (we do not take into account the weak interactions),
unitarity gives 
\be
\label{pole20}
f_h(s) = (1+2 i k T_{\Lambda \Lambda}(k))f^*_h(s)
\ee
where $T_{\Lambda \Lambda} = - 1/(\kappa + ik)$ is the amplitude for
elastic resonance
$\Lambda \Lambda$ scattering, $\kappa = \sqrt{m_\Lambda\eps}$ 
($\eps=2m_\Lambda -m_H$), $s = 4(m^2 + k^2)$, and $k=|{\bf k}|$ 
is the
momentum of the $\Lambda$ hyperon in the c.m.s. Comparing the pole behaviors
of Eqs.~(\ref{pole19}) and~(\ref{pole20}), 
we obtain the expression for the pole of 
$f_h(s)$:
\be
\label{pole21}
f_h(s)=
\frac{\lambda_Hg_{H\Lambda}}{8\kappa(\kappa+ik)}
(1+O(\kappa+ik))\ .
\ee
Substituting~(\ref{pole21}) into~(\ref{eq18}), we find the required
expression for the contribution of the state 
$|\Lambda \Lambda \rangle$ to the spectral
density: 
\be
\label{eq22}
\rho_{\Lambda \Lambda}(s)=\frac{1}{2\pi}
\frac{\lambda_H^2 g_{H\Lambda}^2}{16\kappa^2}
\frac{1}{(s-4 m_\Lambda^2+4\kappa^2)}
\frac{1}{8\pi}\left(\frac{s-4 m_\Lambda^2}{s}\right)\ .
\ee
The interaction constant $g_{H\Lambda}$ can
be found~\cite{Landau} by comparing the pole contribution of the
virtual dihyperon
$H$ to the quantum field-theoretical expression for the amplitude of
elastic $\Lambda \Lambda$ scattering with the pole behavior of the
quantum-mechanical expression for the amplitude of low-energy elastic
resonance $\Lambda \Lambda$ scattering: 
\be
\label{eq23}
g_{H\Lambda}^2=32\pi 4 m_\Lambda \kappa \ .
\ee
It can be seen from Eq.~(\ref{eq22}) that $\rho_{\Lambda \Lambda}(s)$ 
has a resonance peak
near the $\Lambda \Lambda$ threshold with a maximum at the point 
$s_{max} = 4m_\Lambda^2 + 4\kappa^2$ and
a width determined by the value of $\kappa$. Thus, for small  $\kappa$ 
(when the
width of the resonance peak does not exceed the error in the method of
QCD sum rules, i.e. 15\% of $m_H$) this peak and the peak from the
dihyperon $H$ itself fuse in the spectral density $\rho$ into a single
effective peak, the position of whose maximum we determine by fitting
the sum rule. However, since the width of this peak for small  $\kappa$ does
not exceed the error of the method in the determination of $m_H$, by
calculating the position of its maximum we are calculating, within the
accuracy of the method, the value of $m_H$ itself. However, the constant
$\lambda$ in the spectral density~(9) will then no longer be equal to the
projection $\lambda_H$ of the current $h$ onto the dihyperon state 
[see~(2)]. This will be some effective $\lambda$, whose relation of 
$\lambda_H$ we shall now estimate.  

For this, we estimate the contribution of the resonance
peak from the $\Lambda \Lambda$ state to the sum rule in the leading order in
 $\kappa$. Using Eqs.~(\ref{eq22}) and~(\ref{eq23}), we find 
\be
\int_{4m_\Lambda^2}^\infty e^{-s/y}\rho_{\Lambda \Lambda}(s)ds
=e^{-m_H^2/y}\lambda_H^2\frac{1}{\pi}
\frac{\sqrt{m_\Lambda E_0}}{\kappa}\ .
\ee
Here $s_0=(2m_\Lambda + E_0)^2$ must be
taken such that the range of integration spans the resonance peak,
i.e. $E_0$ is approximately equal to the width of the peak. Thus, the
presence of the resonance peak from the $\Lambda \Lambda$ state leads 
to a difference
between the effective constant $\lambda$ in Eq.~(9) and the constant 
$\lambda_H$ for
coupling of the current $h$ to the dihyperon $H$;
\be
\lambda^2=\lambda_H^2(1+\pi^{-1}\sqrt{m_\Lambda E_0}/\kappa) .
\ee
When $\kappa$ is fairly large (i.e. when the width of the $\Lambda \Lambda$
resonance peak is larger than the uncertainty in the method of QCD
sum rules in the determination of the masses), the peak becomes quite
weak and can be included in the effective continuum.  

Finally, we
formulate our main conclusion. The analysis of our sum rules favors
the existence of a dihyperon $H$, in other words, it agrees with Jaffe's
prediction in the framework of the MIT quark-bag model that there
exists a dihyperon $H$ with the quantum numbers of two $\Lambda$
hyperons. However, the accuracy of the method of QCD sum rules, which
in our case is $\sim 20\%$, does not make it possible to determine whether
the mass of the dihyperon $H$ lies above or below the $\Lambda \Lambda$ 
threshold (i.e. whether $H$ is stable).  

The authors are grateful to
A.~N.~Tavkhelidze for interest in the work and to K.~G.~Chetyrkin for
helpful discussions.

\end{document}